\documentclass[aip,jcp,reprint,amsmath,amssymb,floatfix,citeautoscript,nofootinbib]{revtex4-2}
\usepackage{cancel}
\usepackage{amsmath}
\usepackage{mathtools}
\usepackage{physics}
\usepackage{graphicx}
\usepackage{amssymb}
\usepackage{amsthm}
\usepackage{bm}
\usepackage{dcolumn}
\usepackage{longtable}
\usepackage{ragged2e}
\usepackage{newtxtext,newtxmath}
\usepackage[version=3]{mhchem} 
\usepackage[T1]{fontenc}
\usepackage[colorlinks=true,allcolors=blue]{hyperref}
\usepackage[capitalise]{cleveref}
\usepackage{mhchem}

\newcommand{\vE}{{\bm{E}}}
\newcommand{\vP}{{\bm{P}}}
\newcommand{\vmu}{{\bm{\mu}}}
\newcommand{\SN}[1]{{S\textsubscript{N}{#1}}}

\allowdisplaybreaks

\begin{document}
\title{Reaction Rate Theory for Electric Field Catalysis in Solution}

\author{Sohang Kundu}
\affiliation{Department of Chemistry, Columbia University, New York, New York 10027, USA}
\author{Timothy C. Berkelbach}
\email{t.berkelbach@columbia.edu}
\affiliation{Department of Chemistry, Columbia University, New York, New York 10027, USA}
\affiliation{Initiative for Computational Catalysis, Flatiron Institute, New York, New York 10010, USA}

\begin{abstract}
The application of an external, oriented electric field has emerged as an attractive technique for 
manipulating chemical reactions. Because most applications occur in solution, 
a theory of electric field catalysis requires treatment of the solvent, 
whose interaction with both the external field and the reacting species modifies the reaction energetics and
thus the reaction rate.
Here, we formulate such a transition state theory using 
a dielectric continuum model, and we incorporate dynamical effects due to solvent 
motion via Grote-Hynes corrections. We apply our theory to the 
Menshutkin reaction between \ce{CH3I} and pyridine, which is catalyzed by polar solvents, 
and to the symmetric \SN{2} reaction of \ce{F-} with \ce{CH3F}, which is inhibited by polar solvents. 
At low applied field strengths 
when the solvent responds linearly, our theory predicts near-complete 
quenching of electric field catalysis. However, a qualitative treatment of 
the nonlinear response (i.e., dielectric saturation) 
shows that catalysis can be recovered at appreciable 
field strengths as solvent molecules begin to align with the applied field direction. 
The Grote-Hynes corrrection to the rate constant is seen to vary nonmonotonically 
with increasing solvent polarity due to contrasting effects of the screening ability, 
and the longitudinal relaxation time of the solvent.
\end{abstract}

\maketitle

\section{Introduction}

The impact of oriented electric fields on chemical reactions has
been increasingly studied in recent years, especially in the context of
selectivity and catalysis; for an overview, we refer to recent review articles
and perspectives~\cite{Shaik2016,Ciampi2018,Shaik2018,Stuyver2019,Welborn2018}.
The application of an oriented, external electric field to control a chemical reaction
has been clearly demonstrated using
a scanning tunneling microscope break junction~\cite{Aragones2016,Zang2019}, but this
method is hard to scale to industrial levels.
Alternatively, chemical reactions have been influenced using the electric field formed 
spontaneously at the interface of an electrode and an electrolyte~\cite{Gorin2013,Zhang2018}
and the electric field due to functional groups installed on molecular 
catalysts~\cite{Klinska2015,Blyth2019}.

Significant computational work has been done in this field
using gas phase quantum chemistry~\cite{Shetty2020,Shaik2020,Calvaresi2010,BesaluSala2021,Rincon2016,Hoffmann2022,Gopakumar2023}, in conjunction with hybrid quantum mechanics/molecular mechanics~\cite{Lai2010,Talipov2013,Yan2023} as well as 
classical~\cite{Vaissier2017, DuttaDubey2020} and ab initio~\cite{Jiang2023,Wang2017,Cassone2021} molecular dynamics simulations. This has led to important qualitative understanding of the interaction between a molecule's 
electronic structure and an external electric field.
However, there has been considerably less work on the role of solvents and their dynamics, which is an important topic given that scalable methods of electric field catalysis will likely occur in solution.

Consider a molecular system undergoing a chemical reaction from reactant (R)
to transition state (TS) to product (P). Assuming a one-dimensional reaction coordinate,
the transition state theory (TST) approximation~\cite{Eyring1935} to the 
rate constant for barrier crossing is
\begin{equation}
\label{eq:k0}
k_0 = \frac{\omega_\mathrm{R}}{2\pi} e^{-\beta \Delta V^\ddagger}
\end{equation}
where $\Delta V^\ddagger = V_\mathrm{TS} - V_\mathrm{R}$ is the barrier height,
$\omega_\mathrm{R}$ is the vibrational frequency in the reactant well,
and $\beta = 1/k_\mathrm{B}T$.
In the presence of an external electric field $\vE_\mathrm{ext}$,
the energy of the molecular system is changed, 
to lowest order, by $-\vmu\cdot \vE_\mathrm{ext}$
(throughout this work, we neglect the molecular polarizability).

The barrier height is modified,
\begin{equation}
\Delta V^\ddagger(\vE_\mathrm{ext}) = \Delta V^\ddagger - \Delta \vmu^\ddagger \cdot \vE_\mathrm{ext}
\end{equation}
where $\Delta \vmu^\ddagger = \vmu_\mathrm{TS}-\vmu_\mathrm{R}$ is the dipole 
moment difference
between the transition state and the reactant configuration.
The catalytic effect of the external field is then given by the ratio 
\begin{equation}
\label{eq:tst_gas}
k(\vE_\mathrm{ext})/k(0) = \exp\left(\beta \Delta\vmu^\ddagger \cdot \vE_\mathrm{ext}\right),
\end{equation}
where the zero-field rate constant is $k(0) = k_0$ in Eq.~(\ref{eq:k0}).
How large is the above effect? For a dipole moment difference of $|\Delta \vmu^\ddagger| = 10$~D
and an electric field of strength $|\vE| = 1$~V/nm, the energy change is about
5~kcal/mol. Recalling that $k_\mathrm{B}T \approx 0.6$~kcal/mol at 300~K, the change
to the rate is a factor of about $4\times 10^3$, and the rate can be increased or decreased
depending on the relative orientations of the electric field and the dipole moment difference.
For higher values of $|\Delta \vmu^\ddagger|$, the effect is even more pronounced.

For reactions occuring in solution, a polar solvent interacts with both the electric 
field and the molecular system, modifying the thermodynamics and kinetics of the chemical
reaction.
Therefore, this simple picture of electric field catalysis must be modified, which is the
aim of the present work. 
First, we develop a microscopic electrostatic theory to define free energies
for use within adiabatic TST.

Second, we consider the dynamical response of the solvent and calculate dynamical 
corrections to the rate constant within Grote-Hynes theory~\cite{Grote1980}. 
We apply our theory to two \SN{2}-like reactions, the Menshutkin reaction
of pyridine with \ce{CH3I} and a symmetric \SN{2} reaction of \ce{F-} with \ce{CH3F}. 

\begin{figure}
    \includegraphics[scale=1.0]{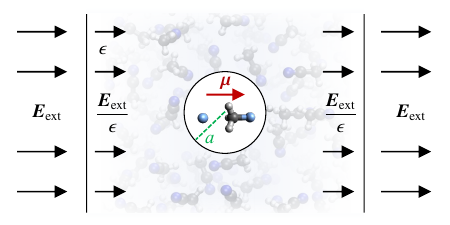}
    \caption[\centering]{
    Dielectric arrangement used in this work to model the electrodynamics of a
molecular complex with dipole moment $\mu$ in a spherical cavity of radius $a$ in a polar 
solvent with dielectric constant $\epsilon$ in an externally applied electric field $\vE_\mathrm{ext}$.
    }
    \label{fig:setup}
\end{figure}

\section{Results and Discussion}
\subsection{Adiabatic Transition State Theory}
To study the influence of a polar solvent, we use the setup shown in Fig.~\ref{fig:setup}.
The molecule is modeled as a dipole $\vmu$ in a spherical cavity of radius $a$. 
The dipole polarizes the surrounding dielectric, which induces a 
constant reaction field~\cite{Onsager1936} inside the cavity,

\begin{equation}
\vE_\mathrm{rxn} = \frac{2}{a^3} \frac{\epsilon-1}{2\epsilon+1} \vmu
\end{equation}
where $\epsilon$ is the 
dielectric constant of the solvent (here and throughout we use atomic units, for which $4\pi\epsilon_0 = 1$).
Simultaneously, an external field $\vE_\mathrm{ext}$ is applied to the dielectric, inside of which 
the field is reduced in magnitude to $\vE_\mathrm{ext}/\epsilon$.

We emphasize that throughout this work, $\vE_\mathrm{ext}$ is the electric field
that would exist in the absence of a solvent.
Due to the formation of bound charges on the surface of the spherical cavity, the 
external electric field generates an electric field inside the cavity,
\begin{equation}
\vE_\mathrm{in} = \frac{3\epsilon}{2\epsilon+1}\frac{\vE_\mathrm{ext}}{\epsilon}
    = \frac{3}{2\epsilon+1} \vE_\mathrm{ext}.
\end{equation}
By the linearity of the Poisson equation, the total field in the cavity is the sum,
\begin{equation}
\vE_\mathrm{tot} = \vE_\mathrm{rxn} + \vE_\mathrm{in},
\end{equation}
leading to the interaction free energy
\begin{equation}
G = -\vmu\cdot\vE_\mathrm{tot} = 
    - \frac{2\mu^2}{a^3} \frac{\epsilon-1}{2\epsilon+1}
    -\frac{3}{2\epsilon+1}\vmu\cdot\vE_\mathrm{ext}.
\end{equation}
Note that the solvation energy, arising from the interaction with the reaction field, is always stabilizing, 
whereas the interaction with the external field depends on orientation.

The TST rate constant in solution is thus
\begin{equation}
\label{eq:tst}
k_\mathrm{TST}(\vE_\mathrm{ext}) 
    = \frac{\omega_\mathrm{R}}{2\pi} e^{-\beta \Delta G^\ddagger(\vE_\mathrm{ext})}
\end{equation}
where
\begin{equation}
\label{eq:dG}
\Delta G^\ddagger(\vE_\mathrm{ext}) = \Delta V^\ddagger 
    - \frac{2(\mu_\mathrm{TS}^2 - \mu_\mathrm{R}^2)}{a^3} \frac{\epsilon-1}{2\epsilon+1}
    -\frac{3}{2\epsilon+1}\Delta\vmu^\ddagger\cdot\vE_\mathrm{ext}.
\end{equation}
Whether the solvation effect is catalytic depends on the sign of
$\mu_\mathrm{TS}^2 - \mu_\mathrm{R}^2$,
which is a well-known effect for chemical reactions in polar solvents.
Clearly, in a nonpolar solvent with $\epsilon =1$, this rate constant reduces 
to Eq.~(\ref{eq:tst_gas}).

The theory so far treats the solvent as a linear dielectric, i.e., in the absence of the cavity,
the solvent polarization is $\vP = (\epsilon-1)\vE_\mathrm{ext}$.
In this approximation, the effect of the reaction field and the external field are separable,
and the catalytic
effect of the external field compared to the reaction in solution is simply determined 
by the ratio
\begin{equation}
\label{eq:tst_E}
k_\mathrm{TST}(\vE_\mathrm{ext})/k_\mathrm{TST}(0) 
    = \exp\left[3 \beta \Delta\vmu^\ddagger\cdot \vE_\mathrm{ext}/(2\epsilon+1)\right].
\end{equation}

Comparison with Eq.~(\ref{eq:tst_gas}) shows that the effect of a polar solvent is
merely to screen the electric field, reducing its strength by about a factor of $\epsilon$.
For common polar solvents with $\epsilon \gtrsim 10$, this would severely limit the
applicability of electric field catalysis in solution.

However, real solvents behave as nonlinear dielectrics, and when they are 
subjected to large electric fields, the polarization saturates with 
increasing field strength due to the near-complete 
alignment of molecular dipole moments. This phenomenon, known as dielectric saturation, 
can be approximately modeled by using a dielectric constant that depends on the strength 
of the applied electric field,
leading to a polarization $\vP = \left[\epsilon(E_\mathrm{ext})-1\right]\vE_\mathrm{ext}$.
A common approximation is Booth's equation~\cite{Booth1951},
\begin{equation}
\label{eq:saturation}
\epsilon(E) = 1 + \frac{3 E_c}{E}\left(\epsilon-1\right)L\left(\frac{E}{E_c}\right),
\end{equation}
where $\epsilon$ is the field-free dielectric constant, $L(x) = \mathrm{coth}(x)-1/x$ 
is the Langevin function, and 

$E_c$ is a microsopic field parameter characteristic of the solvent. 

Within this model, the dielectric constant decreases with increasing field strength
and saturates to $\epsilon = 1$ when $E \gg E_c$. 
Using this field-dependent dielectric constant in Eq.~(\ref{eq:tst_E}) yields 
\begin{equation}
\label{eq:sat_enhancement}
\begin{split}
&k_\mathrm{TST}(\vE_\mathrm{ext})/k_\mathrm{TST}(0) = \exp\left[\frac{3 \beta \Delta\vmu^\ddagger\cdot \vE_\mathrm{ext}}{2\epsilon(E_\mathrm{ext})+1}\right]\\
    &\hspace{1em} \times \exp\left[ \frac{2\beta(\mu_\mathrm{TS}^2 - \mu_\mathrm{R}^2)}{a^3}\left(\frac{\epsilon(E_\mathrm{ext})-1}{2\epsilon(E_\mathrm{ext})+1}
    -\frac{\epsilon-1}{2\epsilon+1}\right)\right],
\end{split}
\end{equation}
which is one of the main results of this work.

\begin{figure}
    \includegraphics[scale=0.4]{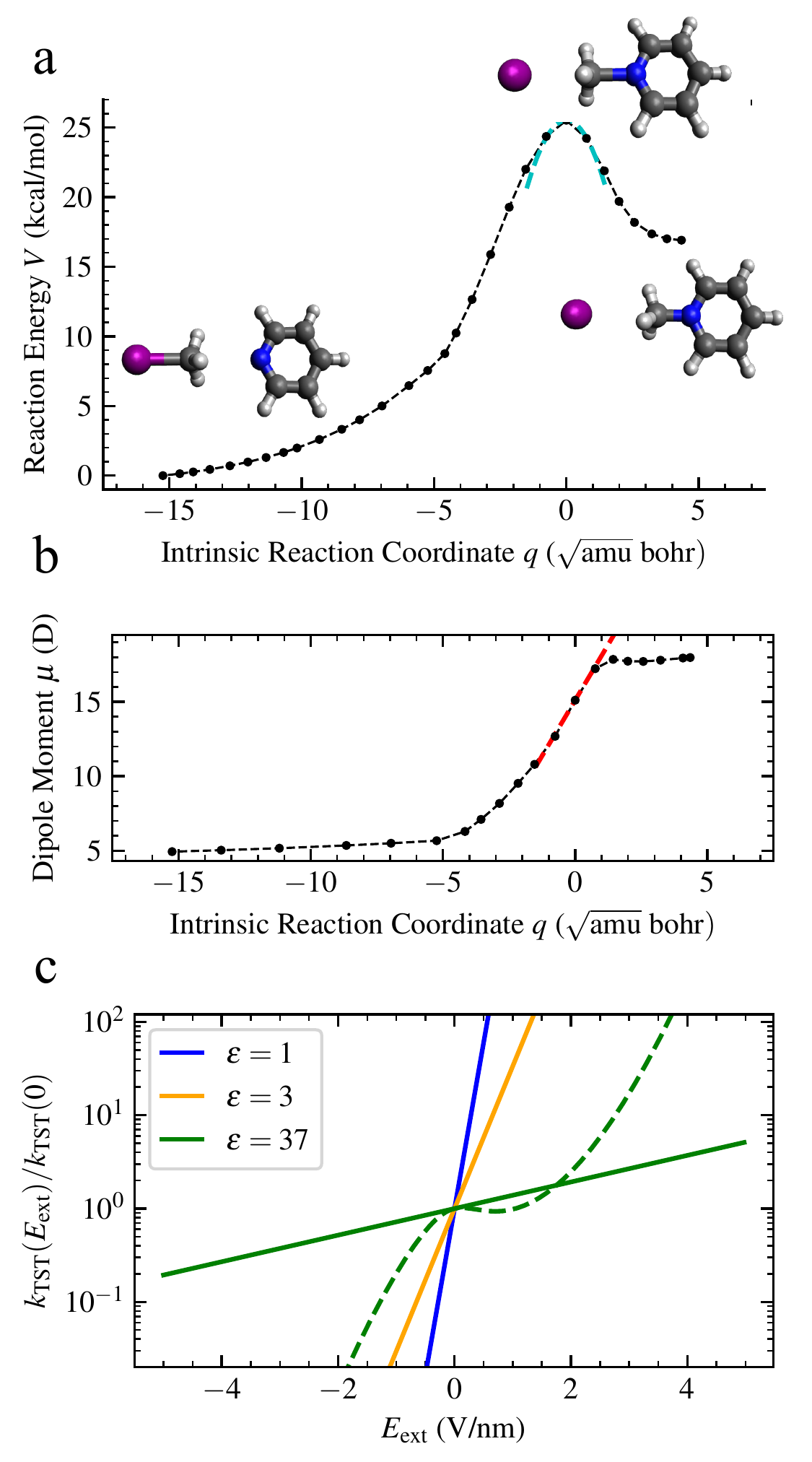}
    \caption[\centering]{
        (a) Reaction energy and (b) dipole moment 
    along the mass-weighted intrinsic reaction coordinate of the Menshutkin  
    reaction between \ce{CH3I} and pyridine. The cyan dashed line in (a) highlights the curvature 
    of the reaction energy at the TS, and the red dashed line in (b) indicates the slope 
    of the dipole moment at the TS. (c) The adiabatic TST rate enhancement due to an external 
    electric field in various solvents with the dielectric constants indicated ($\epsilon=37$ 
    corresponds to acetonitrile). The dashed green line incorporates nonlinear effects of 
    dielectric saturation for acetonitrile.
    }
    \label{fig:Menshcomb}
\end{figure}

The first system to which we apply this theory is the Menshutkin reaction
of pyridine with \ce{CH3I}, which was studied in the same context of
electric field catalysis a few years ago 
by Shaik and co-workers~\cite{DuttaDubey2020} using a combination of
atomistic  molecular dynamics (MD) simulations and quantum chemistry calculations
with implicit and explicit solvent. In the absence of an external electric field, 
the solvent alone was found to have a catalyzing effect on the
reaction (in acetonitrile, the barrier is lowered by 8.5~kcal/mol),
which is a well-known effect~\cite{Sola1991,Castejon1999} from previous experimental
and computational studies. Under an applied electric field, 
polarization of the solvent was found to create a near-complete screeening effect, 
and only when the solvent was maximally polarized beyond a certain external field 
strength (about 1.5~V/nm in acetonitrile), the catalytic effect was recovered, 
leading to an additional reduction of the barrier 
height by 10.6~kcal/mol at $E=5$ V/nm.
We will use these results as a test of our theory, and the numerical comparisons are
summarized in Tab.~\ref{tab:shaik}.

Using the ORCA package~\cite{Neese2017}, we calculated
the minimum energy pathway and dipole moment for the intrinsic reaction 
coordinate~\cite{Fukui1975,Ishida1977} (IRC),
as shown in Fig.~\ref{fig:Menshcomb}a and Fig.~\ref{fig:Menshcomb}b respectively.
Calculations were performed using unrestricted density functional theory (DFT)
with the B3LYP functional~\cite{Becke1988,Becke1993} and the def2-TZVP basis 
set~\cite{Schaefer1994,Weigend2005} (additional details are given in the Supporting 
Information), leading to a barrier height of $\Delta V^\ddagger = 25.5$~kcal/mol.
In the gas phase, this reaction proceeds from neutral reactants to ionic products,
such that the dipole moment is increasing in magnitude throughout the reaction. 
We find  $\mu_\mathrm{R} = 4.9$~D and
$\mu_\mathrm{TS} = 15.1$~D, such that 

$\mu_\mathrm{TS}^2 - \mu_\mathrm{R}^2 > 0$, confirming that the reaction is catalyzed
by a polar solvent.
Based on solvent distribution functions calculated in another recent MD study~\cite{Turan2022},
we estimate a cavity radius of $a = 0.7$~nm.
 
Using these parameters, 
we find that, when no field is applied, the barrier height is lowered by 
8.2~kcal/mol in acetonitrile solvent ($\epsilon=37$).
This is in remarkably good agreement with the value of 8.5~kcal/mol found
by atomistic simulations~\cite{DuttaDubey2020}, indicating an accurate parameterization of
our model. 

In Fig.~\ref{fig:Menshcomb}c, we plot the rate enhancement obtained due to an 
external electric field from Eq.~(\ref{eq:tst_E}), without dielectric saturation.
We consider three values of the solvent dielectric constant,
$\epsilon=1$, 3, and 37. 
As expected, the logarithm of the enhancement is linearly proportional to the field, with a 
slope proportional to $(\epsilon+1)^{-1}$. 
For large external fields and weakly polar solvents, the rate modification can be as much
as $10^5$. However, due to the screening effect of polar solvents, 
the modification is small:
for acetonitrile, with $\epsilon=37$, even an external field of 5~V/nm modifies 
the rate by only a factor of 5.

In the same figure, we also plot the rate enhancement in acetonitrile including
dielectric saturation, i.e., using Eq.~(\ref{eq:sat_enhancement}).  Following
Ref.~\onlinecite{Daniels2017}, we parameterize the Booth model with $E_c
=0.15$~V/nm for acetonitrile.  We see that the rate behavior is more
complicated, and importantly shows a stronger catalytic effect at large fields,
in agreement with the simulations of Shaik and
co-workers~\cite{DuttaDubey2020}. Interestingly, the change in rate for a
saturating dielectric is no longer monotonic and is asymmetric with respect to
the direction of the applied field. 

To understand this change in behavior, let us focus on the
contributions of the reaction field and the external field separately.
Saturation reduces the ability of the solvent to solvate the solute dipole,
reducing the magnitude of the reaction field $\vE_\mathrm{rxn}$ and the solvation energy,
increasing the barrier height.  On the other hand, saturation also reduces the
screening ability of the solvent under applied field, enhancing
$\vE_\mathrm{in}$ compared to its value calculated by neglecting dielectric
saturation. These two effects are in competition when the applied field is
in the positive direction, with the weakening of the reaction field being 
more dominant at low applied field strengths. 
Eventually, the applied field eliminates the screening ability of the solvent,
leading to a recovery of electric field catalysis around $E_\mathrm{ext}=1.5$~V/nm.

\begin{table}[t]
    \centering
    \begin{tabular}{lcccc}
    \hline\hline
              & $\Delta V^\ddagger$ & $\Delta G^\ddagger(0~\mathrm{V/nm})$ & $\Delta G^\ddagger(5~\mathrm{V/nm})$ & $\Delta\Delta G^\ddagger$ \\
    \hline
        Present theory                   & 25.5 & 17.3 &  &  \\
        - Linear dielectric                &      &      & 16.3 & $-1.0$ \\
        - Nonlinear dielectric             &      &      & 11.9 & $-5.4$ \\
        - With electrofreezing             &      &      &  8.7 & $-8.6$ \\
        Simulation [\onlinecite{DuttaDubey2020}] & 27.4 & 18.9 &  8.3 & $-10.6$ \\
    \hline\hline
    \end{tabular}
    \caption{Comparison of the gas phase barrier $\Delta V^\ddagger$ and solution phase 
        barrier $\Delta G^\ddagger$, without and with an applied electric field, between the present theory
        and the atomistic simulation results of Ref.~\onlinecite{DuttaDubey2020} for the
        Menshutkin reaction in acetonitrile solvent.
        The change to the solution phase barrier due to the electric field is denoted 
        $\Delta\Delta G^\ddagger \equiv \Delta G^\ddagger(E) - \Delta G^\ddagger(0)$.
        Theoretical predictions are presented for the nonlinear theory that includes dielectric saturation,
        and the predictions of the linear theory are given in parentheses.
        All energies are in kcal/mol.
    }
    \label{tab:shaik}
    \end{table}

Focusing on the case with $E_\mathrm{ext}=5$~V/nm, we find that the field reduces the barrier
height by 1.0~kcal/mol when we neglect dielectric saturation but by 5.4~kcal/mol
when we include it. This latter value is smaller than the lowering
of 10.6~kcal/mol observed in simulations, but clearly a significant improvement,
especially considering the simplicity of the model.
Our Booth fit in Eq.~(\ref{eq:saturation}) gives 
$\epsilon(E_\mathrm{ext}=5~\mathrm{V/nm}) = 4.2$, but MD simulations reported in
Ref.~\onlinecite{Daniels2017} indicated $\epsilon(E_\mathrm{ext}=5~\mathrm{V/nm}) \approx 2.4$,
which when used in our theory predicts a barrier height reduction of 8.6~kcal/mol, 
in much better agreement with the results of Shaik and co-workers~\cite{DuttaDubey2020}.
However, as pointed out in Ref.~\onlinecite{Daniels2017}, this lower dielectric constant is
due to an electrofreezing transition in acetonitrile that occurs around 
$E_\mathrm{ext} \approx 3~\mathrm{V/nm}$ at room temperature.
This important effect must be kept in mind when considering the application of
strong electric fields to the catalysis of reactions in polar solvents.

\begin{figure}[t]
    \includegraphics[scale=0.4]{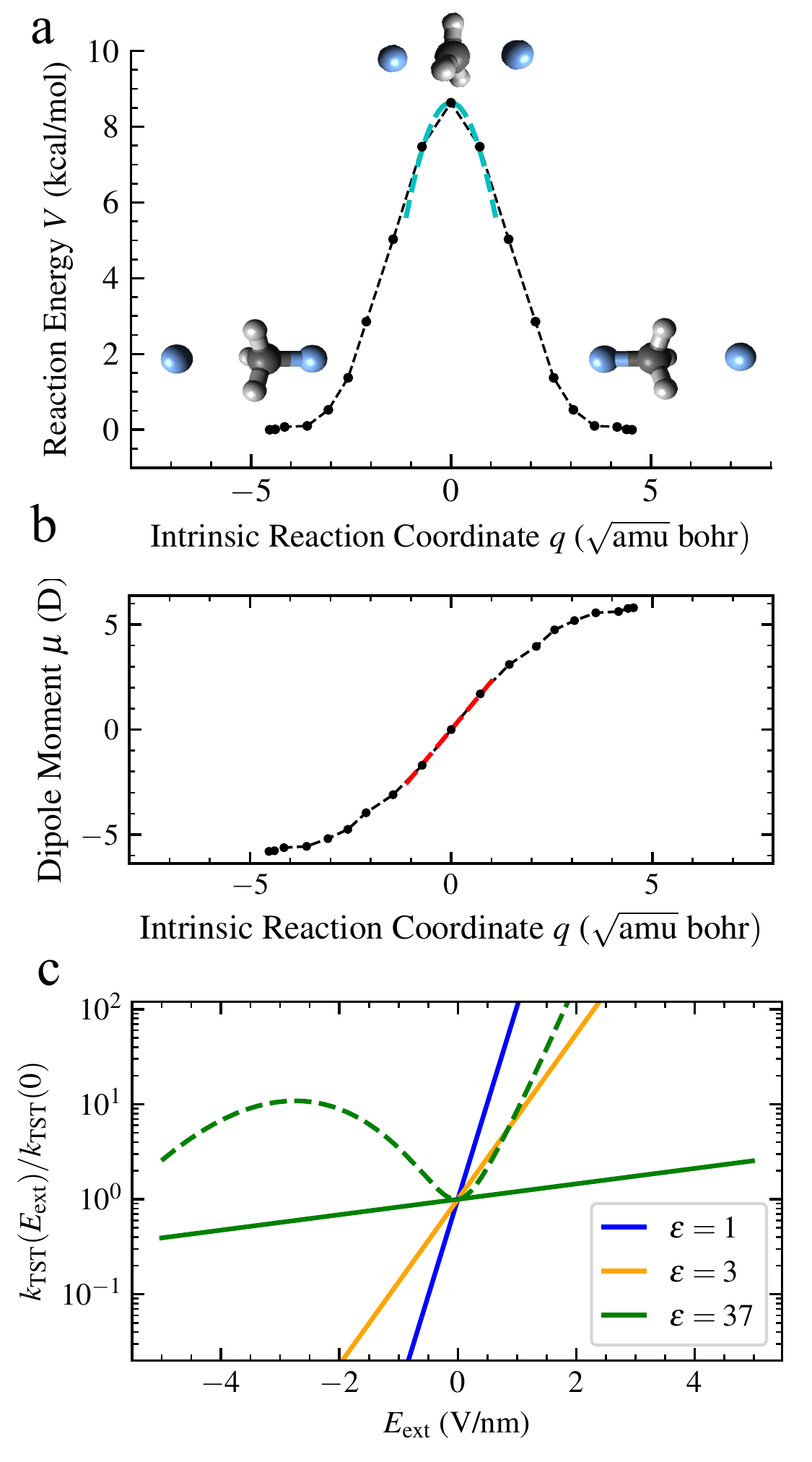}
    \caption[\centering]{Same as Fig.~\ref{fig:Menshcomb} but for 
    the symmetric \SN{2} reaction between \ce{CH3F} and \ce{F-}.} 
    \label{fig:CH3Fcomb}
\end{figure}

Next, we apply the theory to the symmetric S$_\mathrm{N}$2 reaction of \ce{F-} with
\ce{CH3F}. The results are shown in Fig.~\ref{fig:CH3Fcomb}.
Using the same computational methods, we calculate a reaction barrier of
$\Delta V^\ddagger = 8.6$~kcal/mol and dipole moments of $\mu_\mathrm{R}=-5.8$~D, 
$\mu_\mathrm{TS}=0$, such that $\mu_\mathrm{TS}^2 - \mu_\mathrm{R}^2 < 0$ and a polar
solvent inhibits the reaction.
For example, estimating $a=0.3$~nm
we find that the barrier height increases by 17.3~kcal/mol in acetonitrile. 

Without accounting for dielectric saturation, 
an applied electric field is strongly screened by the polar solvent, as was observed
for the Menshutkin reaction.
However, significant differences emerge when we include dielectric saturation. 
Because the solvent reaction field leads to an increase in the barrier height, 
we find that the field-induced inhibition of solvation of the molecular dipole 
leads to a catalytic effect at low applied field strengths independent of the 
direction of the field. 
To the best of our knowledge, this effect---inhibitation of reactant stabilization---is 
a new mode for electric field catalysis, and it would be simple to implement due to
its independence of the field direction.
For stronger fields, we see that the catalytic effect
continues to grow when the field is positive, but reverses its behavior when the
field is negative, because the external field and reaction field are now in competition. 

This completes our development of an adiabatic TST of electric field catalysis
with nonlinear dielectric effects. However, it is known that the dynamics of polar
solvents can significantly impact the rate of reactions, especially those whose
charge density changes significantly when crossing the barrier.
In the next section, we consider these dynamical corrections using Grote-Hynes
theory in the presence of an external electric field.

\subsection{Dynamical Corrections}

Within TST, the reaction is assumed to be adiabatic, i.e., the solvent responds instantaneously to
changes in the molecular reaction dynamics. In reality, the solvent response lags, and recrossing
events are responsible for a reduction in the rate constant compared to its TST value.
Here, we incorporate dynamical corrections to the rate constant,
\begin{equation}
    k_\mathrm{GH} = \kappa k_\mathrm{TST},
\end{equation}
where $\kappa$ is the transmission factor calculated with Grote-Hynes theory~\cite{Grote1980}. 
We aim to understand how $\kappa$ depends on the solvent polarity and the 
external electric field.

Specifically, we
consider the dynamics of the IRC $q$ near the TS, which we
take to occur at $q=0$. In this region, we treat the energy barrier as parabolic, and
we linearize the transition state dipole moment $\vmu_\mathrm{TS}(q) \approx \vmu' q$,
as shown in Figs.~\ref{fig:Menshcomb} and \ref{fig:CH3Fcomb},
leading to the free energy surface~\cite{Haynes1993}

\begin{equation}
\begin{split}
\Delta G(q) &= \Delta V^\ddagger - \frac{1}{2}\omega_\mathrm{b,0}^2 q^2
    - \frac{2|\mu'|^2}{a^3} \frac{\epsilon-1}{2\epsilon+1} q^2  \\
    &\hspace{1em} - \frac{3}{2\epsilon+1}\vmu'\cdot\vE_\mathrm{ext} q
    \\
&\equiv \Delta V^\ddagger - \frac{1}{2}\omega_\mathrm{b,eq}^2 q^2 - F_\mathrm{ext} q,
\end{split}
\end{equation}

where $\omega_\mathrm{b,0}$ is the barrier frequency in the gas phase,
$\omega_{\mathrm{b,eq}} = \sqrt{\omega_\mathrm{b,0}^2 + 2\Lambda}$ is the 
(adiabatic) free energy barrier
frequency,
and
\begin{equation}
\label{eq:lambda}
\Lambda = \frac{2|\mu'|^2}{a^3}\frac{\epsilon-1}{2\epsilon + 1}
\end{equation}
is the solvent reorganization energy (see below).
In this limit, the dynamics is described by the generalized Langevin equation
\begin{equation}
\ddot{q} = - \omega_\mathrm{b,eq}^2 q - \int_0^t dt' \zeta(t-t') \dot{q}(t') + \delta F_q(t) + F_\mathrm{ext},
\end{equation}
where $\zeta(t)$ is a friction kernel due the interaction between the molecular dipole and the solvent,
and $\delta F_q(t)$ is a random force satisfying the fluctuation-dissipation relation
$\beta \langle \delta F_q(t) \delta F_q(0) \rangle = \zeta(t)$.
The Grote-Hynes correction is
\begin{equation}    
\kappa = \frac{\lambda^\ddagger}{\omega_\mathrm{b,eq}}, \quad{}
\lambda^\ddagger = \frac{\omega_\mathrm{b,eq}^2}{\lambda^\ddagger 
    + \hat{\zeta}(\lambda^\ddagger)},
\end{equation}
where the reactive frequency $\lambda^\ddagger$ is the lowest positive root of
the transcendental equation on the right.

As a model of the solvent dynamics, we use the Debye approximation to the 
dielectric function~\cite{Nee1970,Bagchi1984,VanderZwan1985,Moro1989},
\begin{equation}
\hat{\epsilon}(z) = 1 + \frac{\epsilon-1}{1+z\tau_\mathrm{D}},
\end{equation}
where $\epsilon$ is the static dielectric constant, $\tau_\mathrm{D}$ is the
Debye relaxation time, and $\hat{f}(z)$ indicates the Laplace transform.
Then, as shown in the Supporting Information, 
the friction kernel is given simply by
\begin{equation}
    \hat{\zeta}(z) = \frac{\Lambda}{z+\tau_\mathrm{L}^{-1}},
\end{equation}
where $\Lambda$ is the reorganization energy in Eq.~(\ref{eq:lambda}), and 
$\tau_\mathrm{L} = 3\tau_\mathrm{D}/(2\epsilon+1)$ is
the longitudinal relaxation time. 
In the time domain, this model yields simple exponential
decay of the friction kernel, $\zeta(t) = \Lambda e^{-t/\tau_\mathrm{L}}$.

\begin{figure}[t]
    \includegraphics[scale=0.43]{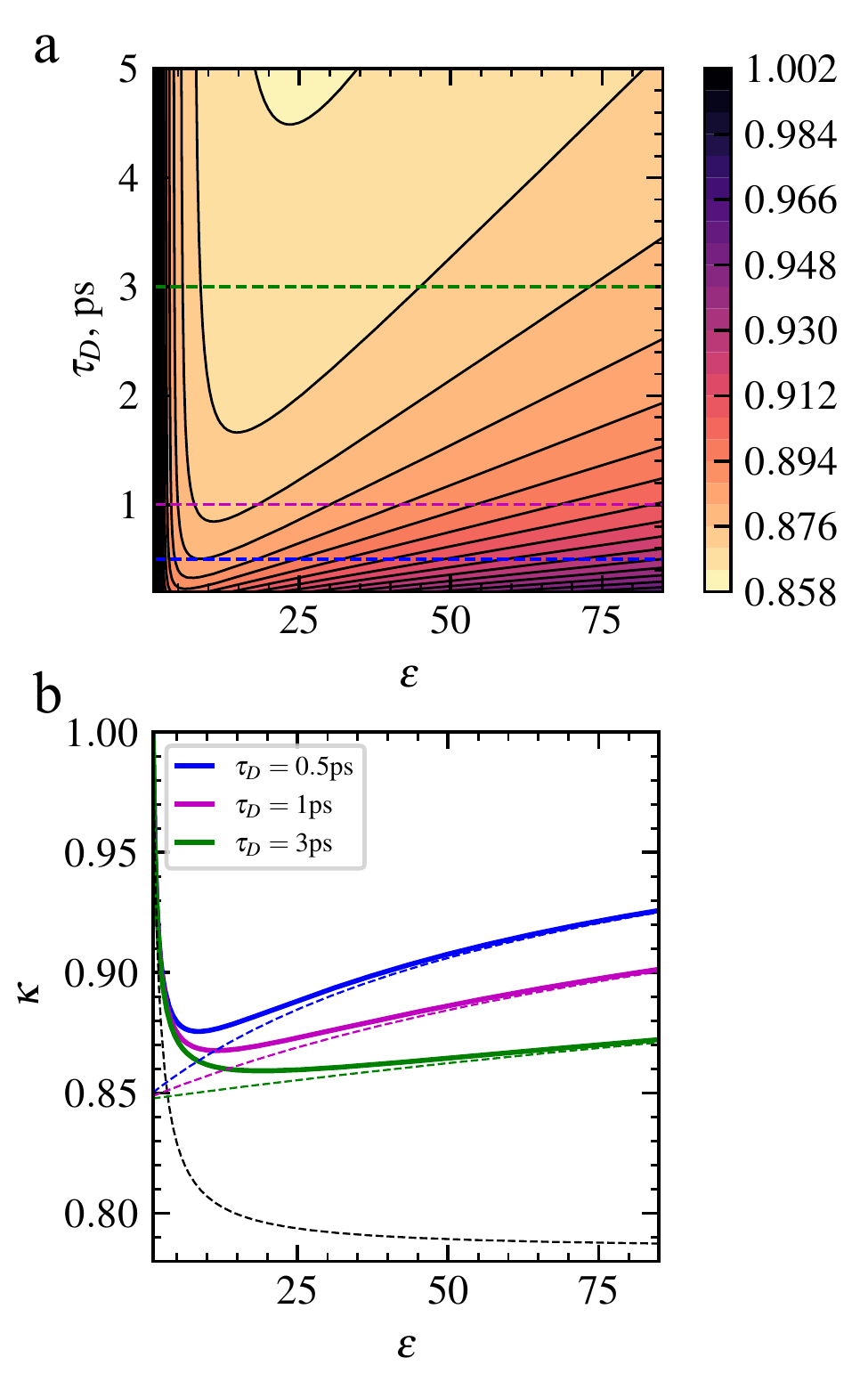}
    \caption[\centering]{Panel (a) shows the Grote-Hynes transmission coefficient $\kappa$ as a function of the solvent's Debye relaxation time
    $\tau_\mathrm{D}$ and static dielectric constant $\epsilon$. The blue, magenta and green lines (bottom to top) are slices 
    at $\tau_\mathrm{D} = 0.5$, 1, and 3~ps, results for which are shown in panel (b). For acetonitrile, 
    $\tau_D\approx3$~ps at room temperature. Also shown are the transmission coefficients obtained by artifically keeping $\tau_{L}$ constant (black dashed line) and 
    $\Lambda$ constant (colored dashed lines).} 
    \label{fig:GH-kappa-2D}
\end{figure}

\begin{figure}
    \includegraphics[scale=1]{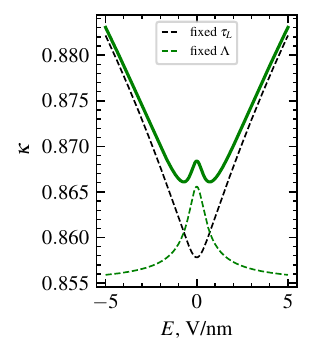}
    \caption[\centering]{The effect of dielectric saturation (nonlinearities) on the Grote-Hynes correction $\kappa$
    for the symmetric S$_\mathrm{N}$2 reaction in acetonitrile. 
    As in Fig. \ref{fig:GH-kappa-2D}, we also show results obtained 
    when $\tau_{\mathrm{L}}$ and $\Lambda$ are separately held constant.} 
    \label{fig:GH-nonlinear}
 \end{figure}

Within a linear dielectric theory, a static external electric field does not
modify the dynamics of equilibrium solvent fluctuations, and therefore the GH
transmission coefficient is independent of the external field. However, if we
allow the static dielectric constant (which enters the Debye model) to be
field dependent owing to dielectric saturation, then the equilibrium 
solvent dynamics also acquire a field dependence.

We only present results for the symmetric S$_\mathrm{N}$2 reaction, which we found
to exhibit the largest dynamical corrections, mostly due to its smaller cavity radius $a$ 
(analogous results for the Menshutkin reaction are presented in the Supporting Information). 
For both reactions, we
always find $\kappa > 0.8$, indicating that TST is a good approximation.
From the IRC calculations of the S$_\mathrm{N}$2 reaction, we extract 
$\mu^\prime=0.021$~D~$\mathrm{Bohr^{-1}}$~$\mathrm{amu^{-0.5}}$ and
$\omega_\mathrm{b,0}=450.4$~cm$^{-1}$.

In Fig.~\ref{fig:GH-kappa-2D}, we plot the Grote-Hynes transmission coefficient $\kappa$ as a function of
the Debye relaxation time $\tau_\mathrm{D}$ and static dielectric constant $\epsilon$. Interestingly, for fixed $\tau_\mathrm{D}$, 
we see non-monotonic behavior as a function of $\epsilon$. Results at three example values of $\tau_\mathrm{D} = 0.5$, 1, 
and 3~ps are shown in panel (b), where $\tau_\mathrm{D} \approx3$~ps is representative of acetonitrile at room temperature~\cite{Stoppa2015}. 
Such non-monotonic behavior is usually associated with a 
resonance between the solvent response function and the barrier frequency~\cite{Li2021}. However, as we show in 
Fig.~\ref{fig:GH-kappa-2D}(b), the minimum in $\kappa$ occurs at a much lower frequency than that of the reaction barrier. 

The non-monotonic behavior arises instead because both the reorganization energy $\Lambda$ and the longitudinal relaxation
time $\tau_\mathrm{L}$ depend on $\epsilon$, with opposite effects on the transmission coefficient. With increasing $\epsilon$,
the reorganization energy increases from 0 to $|\mu'|^2/a^3$, reflecting stronger coupling between the dipole and the solvent, 
which increases the probability of recrossing and thus lowers $\kappa$. However, with increasing $\epsilon$, 
the longitudinal relaxation time $\tau_\mathrm{L}$
decreases from $3\tau_\mathrm{D}$ to 0, reflecting a more adiabatic solvent response, which increases $\kappa$.
It is the competition of these two effects that yields the non-monotonic behavior. To observe the two effects in isolation,
we also show results obtained by artificially keeping $\tau_\mathrm{L}$ constant (dashed black line), and $\Lambda$ constant
(colored dashed lines) in Fig.~$\ref{fig:GH-kappa-2D}$. 

Finally, we consider the effect of dielectric saturation on solvent dynamics. As mentioned before, the 
Grote-Hynes transmission coefficient only depends on the applied electric field through the modified dielectric constant. 
In Fig.~\ref{fig:GH-nonlinear}, we plot transmisison coefficients obtained for different applied field strengths, specifically for 
the case of acetonitrile with zero-field values $\epsilon = 37$ and $\tau_\mathrm{D} = 3$~ps (we assume
that $\tau_\mathrm{D}$ is unaffected by the applied field strength, but this assumption would
 have to be tested). 
With increasing field, $\epsilon$ is reduced and, since $\kappa$ shows a nonmonotonic dependence on $\epsilon$,
a similar variation is also observed with applied field strength. 
As in Fig. \ref{fig:GH-kappa-2D}, 
the two competing effects responsible for this nonmonotonic dependence 
arise from the simultaneous modulation of the longitudinal relaxation time
$\tau_{\mathrm{L}}$ and the reorganization energy $\Lambda$.

\section{Conclusions}
We have developed a theory for reaction rates in polar solvents in 
the presence of an externally applied electric field parameterized by only a few
physical values.
Importantly, we accounted for dielectric saturation, i.e., the reduction of the
solvent dielectric constant in the presence of an external electric field.
In the TST approximation, we used our theory to study the
Menshutkin methyl transfer reaction between \ce{CH3I} and pyridine 
and the \SN{2} reaction between  \ce{CH3F} and \ce{F-}.
Despite its simplicity, we find that our theory is in nearly quantitative agreement 
with results obtained using fully atomistic simulations~\cite{DuttaDubey2020}.
We therefore expect that our theory will be useful to experimentalists and others
interested in rapid predictions without the cost of atomistic simulations.
Our results confirm the recovery of electric field catalysis
at large field strengths due to dielectric saturation, 
but we warn about the possibility of electrofreezing of the solvent.

Finally, we relax the adiabatic approximation by computing Grote-Hynes dynamical corrections 
to the rate constant within the Debye model for the solvent.
We considered dynamical corrections to TST via Grote-Hynes theory, but 
for the two reactions studied, we find that dynamical corrections reduce the TST rate by
only about 20\%.
We observe an interesting nonmonotonic dependence of the correction factor 
on the electric field strength, due to the opposite ways in which the solvent
reorganization energy and longitudinal relaxation time depend on the field through
the dielectric constant.
However, the scale of the variations in the correction factor is only about 2\%
over an accessible range of electric field strengths.

Despite the successes of our theory, it has several limitations.
We have neglected polarizability and higher order responses,
and we have described the solvent using simple dielectric continuum theory.
Although we employed simple models for the dielectric constant's 
saturating behavior and dynamical response, these can be easily replaced
by more accurate parameterizations or experimental data.
Finally, we have ignored the rotational dynamics of the reacting complex.
It would be useful to understand the impact of all of these approximations in cases
where our theory fails to agree with experiment. 

\vspace{1em}

\section*{Acknowledgements}
T.C.B.~thanks David Limmer and Latha Venkataraman for helpful discussions.
This work was supported by the Columbia Center for Computational Electrochemistry. 
We acknowledge computing resources from Columbia University’s Shared Research Computing Facility project, which is supported by NIH Research Facility Improvement Grant 1G20RR030893-01, and associated funds from the New York State Empire State Development, Division of Science Technology and Innovation (NYSTAR) Contract C090171, both awarded April 15, 2010.

\newpage
\widetext
\begin{center}
    \textbf{Supporting Information for: Reaction Rate Theory for Electric Field Catalysis in Solution}
\end{center}
\setcounter{section}{0}
\setcounter{equation}{0}
\setcounter{figure}{0}
\setcounter{table}{0}
\setcounter{page}{1}
\makeatletter
\renewcommand{\theequation}{S\arabic{equation}}
\renewcommand{\thesection}{S\arabic{section}}
\renewcommand{\thefigure}{S\arabic{figure}}

\section{Computational Details}
\label{app:compute}
All unrestricted density functional theory (DFT) calculations were performed with the ORCA
quantum chemistry package~\cite{Neese2017} using the B3LYP functional~\cite{Becke1988,Becke1988} and 
the def2-TZVP~\cite{Schaefer1994,Weigend2005} basis set. 
For both reactions, the transition state (TS) geometry was first optimized starting with 
a reasonable guess structure. Next, an intrinsic reaction coordinate~\cite{Fukui1975,Ishida1977} 
(IRC) calculation 
was performed using the optimized TS to obtain the reaction energy profile. 
The IRCs were converged with a maximum gradient threshold of $1\times10^{-3}$ Hartree/Bohr 
and a root-mean-square gradient threshold of $2\times10^{-4}$ Hartree/Bohr.
Finally, for each geometry obtained from the IRC calculation, a separate single point calculation
was performed at the same level of theory to extract the dipole moment. 

\section{Derivation of the friction kernel}
\label{app:gle}

Here we derive the GLE friction kernel relevant for the interaction between a continuum polar solvent 
and a dipole whose magnitude can change with time.
The friction kernel $\zeta(t)$ is related to the fluctuating random force
$\delta F_q(t)$ by the fluctuation-dissipation relation
\begin{equation}
\label{eq:friction}
\zeta(t) = \beta \langle \delta F_q(t) \delta F_q(0)\rangle 
= \frac{|\mu'|^2}{3k_\mathrm{B}T}\langle \vE_\mathrm{rxn}(t)\cdot\vE_\mathrm{rxn}(0)\rangle
\end{equation}
where $\langle \vE_\mathrm{rxn}(t)\cdot\vE_\mathrm{rxn}(0)\rangle$ is the equilibrium correlation function of the fluctuations
of the reaction field. The dynamics of the reaction field follow those of the dipole moment via
\begin{equation}
\vE_\mathrm{rxn}(t) = \int_{-\infty}^t dt' \chi(t-t')\vmu(t').
\end{equation}
where the response function is
\begin{equation}
\label{eq:response}
\chi(t) = -\frac{\beta}{3}\frac{d}{dt} \langle \vE_\mathrm{rxn}(t)\cdot\vE_\mathrm{rxn}(0) \rangle
\end{equation}
Combining Eqs.~(\ref{eq:friction}) and (\ref{eq:response}) relates the friction kernel to the response
function,
\begin{equation}
\frac{d\zeta(t)}{dt} = -|\mu'|^2 \chi(t)
\quad{}\mathrm{or}\quad{}
z\hat{\zeta}(z) - \Lambda = -|\mu'|^2 \hat{\chi}(z),
\end{equation}
where we have taken the Laplace transform, and $\Lambda \equiv \zeta(t=0)$ is the reorganization energy.
Within our model of a dipole in a spherical cavity, a boundary value problem yields the response function
\begin{equation}
\hat{\chi}(z) 
    = \frac{2}{a^3} \frac{\hat{\epsilon}(z)-1}{2\hat{\epsilon}(z)+1}.
\end{equation}
Using the Debye approximation for the dielectric function presented in the text gives the
friction kernel
\begin{equation}
\hat{\zeta}(z) = \frac{\Lambda}{z+\tau_\mathrm{L}^{-1}}
\quad{}\mathrm{or}\quad{}
\zeta(t) = \Lambda e^{-t/\tau_\mathrm{L}},
\quad{}\mathrm{with}\quad{}
\Lambda = \frac{2|\mu'|^2}{a^3}\frac{\epsilon-1}{2\epsilon+1}.
\end{equation}

\section{Dynamical Correction Factor for the Menshutkin Reaction}
\label{app:Mensh-kappa}
\begin{figure}
    \includegraphics[scale=0.43]{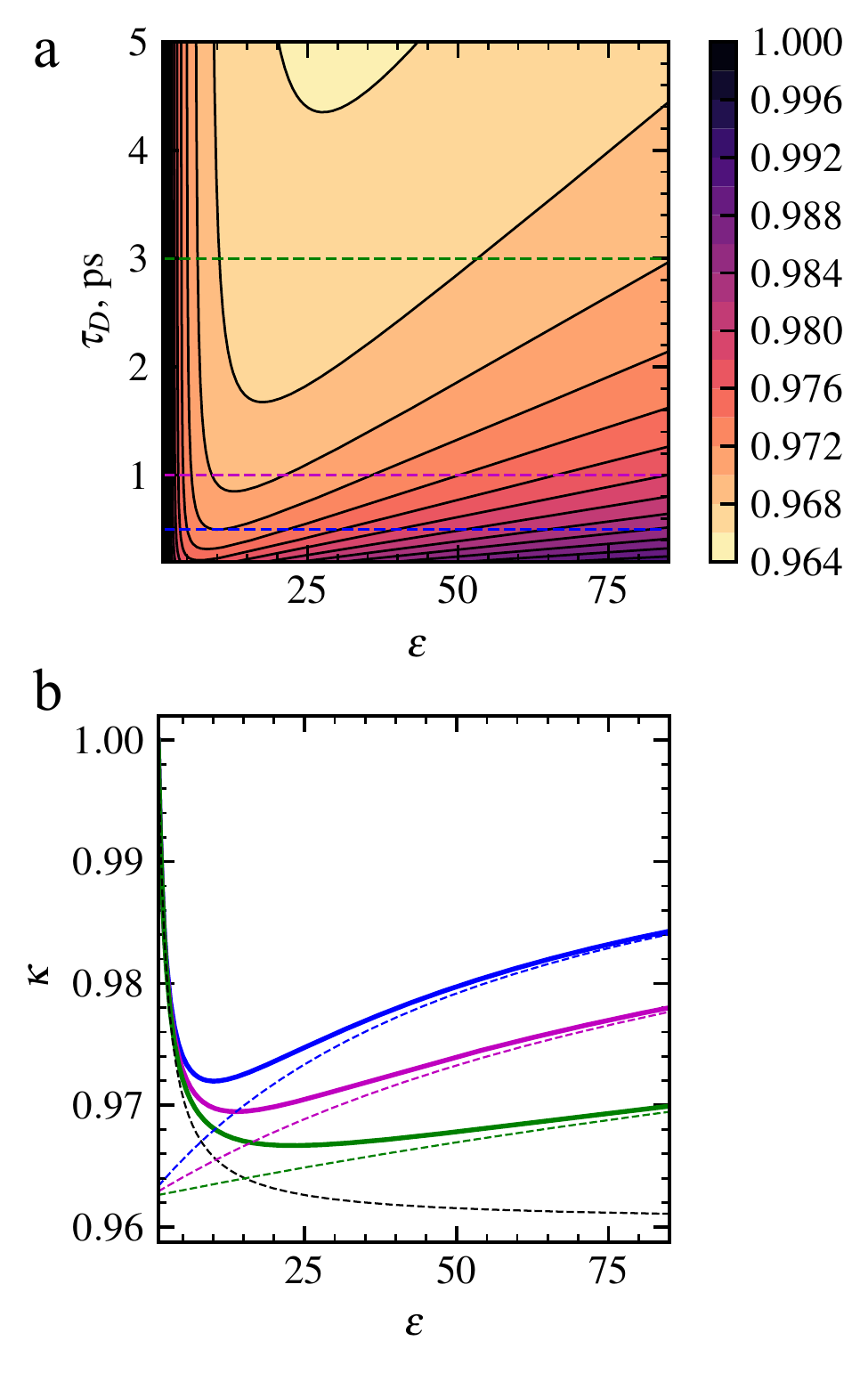}
    \caption[\centering]{The Grote-Hynes transmission coefficient $\kappa$ as a function of the solvent's Debye relaxation time
    $\tau_\mathrm{D}$ and static dielectric constant $\epsilon$ for the Menshutkin reaction (panel (a)). The blue, magenta and green lines (bottom to top) are slices 
    at $\tau_\mathrm{D} = 0.5$, 1, and 3~ps, results for which are shown in panel (b).} 
    \label{fig:Mensh-GH-kappa-2D}
\end{figure}

 Since the Grote-Hynes correction to the transition state theory (TST) rate constant was found to be 
 qualitatively identical for the two reactions, only results for the S$\mathrm{_N}$2 reaction were shown in 
 the main text. We provide the analogous figures for the Menshutkin reaction in 
 Fig. \ref{fig:Mensh-GH-kappa-2D}, for comparison.

\end{document}